\documentclass{sigkddExp}
\begin{document}
\title{Crime in Urban Areas:\\A Data Mining Perspective}
\numberofauthors{1}

\author{
\alignauthor Xiangyu Zhao and Jiliang Tang \\
       \affaddr{Data Science and Engineering Lab, Michigan State University} \\
       \email{\{zhaoxi35, tangjili\}@msu.edu}
}
\date{30 July 1999}
\maketitle

\begin{abstract}
Urban safety and security play a crucial role in improving life quality of citizen and the sustainable development of urban. Traditional urban crime research focused on leveraging demographic data, which is insufficient to capture the complexity and dynamics of urban crimes. In the era of big data, we have witnessed advanced ways to collect and integrate fine-grained urban, mobile, and public service data that contains various crime-related sources as well as rich environmental and social information.  The availability of big urban data provides unprecedented opportunities, which enable us to conduct advanced urban crime research. Meanwhile,  environmental and social crime theories from criminology provide better understandings about the behaviors of offenders and complex patterns of crime in urban. They can not only help bridge the gap from what we have (big urban data) to what we want to understand about urban crime (urban crime analysis); but also guide us to build computational models for crime. In this article, we give an overview to key theories from criminology, summarize crime analysis on urban data, review state-of-the-art algorithms for various types of computational crime tasks and discuss some appealing research directions that can bring the urban crime research into a new frontier.
\end{abstract}

\section{Introduction}
\label{sec:introduction}

We are living in a rapidly urbanizing world. The United Nations predicted that by 2050 about 64\% of the developing world and 86\% of the developed world will be urbanized, which means that the urban population will be bigger than the world population today. Nowadays, the relationship between urbanization and inequalities of urbanites (such as education level and wealth gap) has been extensively studied~\cite{behrens2014survival,oyvat2010urbanization}, and a good amount of research suggests that places with inequalities are more likely to have high crime rates~\cite{crawford2017crime,braithwaite2013inequality,hagan1995crime}. For instance, over the period of 1980 to 2000, recorded crimes increased from 2300 to 3000 for every 100,000 people~\cite{un2012enhancing}. Recent studies have shown that urban safety is closely related to the quality of citizen's life and the sustainable development of urban~\cite{couch2000urban}.  Safety is one of the most fundamental physical and psychological needs of residents. Meanwhile, sustainable urban development will only be achieved when well-planned city-wide, gender-sensitive, community-based, integrated and comprehensive urban crime prevention and safety strategies have been put in place. 

Traditional urban crime research is mainly based on conventional demographic data, i.e., statistical socioeconomic characteristics of a population, such as education level~\cite{ehrlich1975relation}, income level and wealth gap~\cite{kennedy1998social,patterson1991poverty}, and ethnic and religious difference~\cite{braithwaite1989crime}. However, demographic data is insufficient to understand the dynamics and complexity of crimes. First, most demographic features are relatively stable over an extended period of time, which cannot capture the dynamic nature within a specific community. Second, the vast majority of communities in the urban share similar demographic features, thus it becomes difficult to capture the differences between different communities~\cite{wang2016crime}. Recently, with the massive development of new techniques for fine-grained data collection and integration, numerous urban crime-related data has been recorded, which provides sources that contain helpful context information about urban crime. For instance, human mobility provides useful environmental factors such as the function of a region and residential stability, which can significantly impact criminal activities according to environmental criminology~\cite{brantingham1981environmental}; while meteorological data like weather information has been proven to be related to urban crimes~\cite{cohn1990weather,ranson2014crime}. Thus, big urban data contains rich and fine-grained context information about when and where the data is collected. Such information not only enables us to understand the dynamics of crime such as how crime evolves; but also allows us to study crime from various perspectives.  Hence, big urban data provides us unprecedented opportunities to conduct advanced investigations on urban crime. 

On the one hand, there are many criminal theories developed from criminology to explain various types of criminal phenomena. For instance, according to environmental criminology such as routine activity theory~\cite{cohen1979social} and rational choice theory~\cite{cornish2014reasoning}, crime distribution is highly determined by time and space, and environmental factors offered by human mobility can significantly affect criminal activities. Meanwhile, social criminology such as social disorganization theory directly links crime rates to neighborhood ecological characteristics~\cite{shaw1942juvenile}, while culture conflict theory suggests that the root cause of criminality can be found in a clash of values between differently socialized groups over what is acceptable or proper behavior. These theories explore many aspects of criminal behaviors and are crucial to understand and explain how and why crime occurs. Therefore, they can help bridge the gap from what we have (big urban safety data) to what we want to understand about urban crimes (urban crime analysis). 

On the other hand, urban data is typically large-scale, noisy, dynamic and heterogeneous, thus efficient and effective computational solutions are desired.  In order to facilitate crime research with big urban data, a variety of computational tasks have been proposed, which have encouraged a large body of computational models by integrating criminal theories as well as crime patterns.  It is timely and necessary to provide an overview about urban crime from a data mining perspective. Then, the remaining of the article is organized as follows. In Section 2, we introduce environmental and social criminal theories from criminology. Then we summarize the major urban crime patterns in Section 3. In Section 4, we introduce key computational crime tasks with representative algorithms. Finally, we conclude the work with discussions on possible research directions.

\section{Criminology}
\label{sec:criminology}
Crime is a complicated and multidimensional event occurring when the law, offender and target (person or object) converge within time and place~\cite{brantingham1981afterword}. Understanding the offenders' behavior and crime patterns plays an essential role in understanding crime. Consequently, it is beneficial to be acquainted with the theories from criminology.

\subsection{Environmental Criminal Theories}
Environmental criminology focuses on criminal patterns within particularly built environment and analyzes the impacts of the external variables on people's emotional behavior. These consist of space (geography), time, law, and offender, in addition to target or victim. 

\subsubsection{Routine Activity Theory}
Routine activity theory explains crime in terms of crime opportunities that happen in daily life~\cite{felson1994crime,cohen1979social}. Three elements should converge in time and space for a crime opportunity, i.e.,  a motivated offender, a suitable target or victim, and the absence of a capable guardian. A guardian at a place, like a street, could contain security guards or even ordinary pedestrians who would witness the criminal act and possibly intervene or report it to police.  This theory is expanded by the addition of the fourth element of ``place manager" who has the capacity to take nuisance abatement measures~\cite{eck1997reducing}.

\subsubsection{Crime Pattern Theory}
Crime Pattern Theory is to understand why crimes are committed in particular areas. Crime is not really random -- it is either planned or opportunistic. Based on the theory, crime happens when the activity space of the victim or target intersects with that of an offender. Crime Pattern Theory has three main notions -- node, path and edge~\cite{clarke1998opportunity}. Node is an specific area of activity that an individual uses frequently. Path is the route that the individual takes to and from typical areas of activity in everyday life.  Edges are the boundaries of an individual's awareness space. 

\subsubsection{Rational Choice Theory}
Rational choice theory aims to help in thinking about situational crime prevention~\cite{clarke1993introduction}. The assumption is that crime is purposive behavior made to fulfill the criminal's needs for things such as money, status and excitement. Meeting these requirements involves the making of decisions and choices as these are constrained by limits, ability, and the accessibility of relevant information~\cite{clarke1997situational}. For instance, if telephone or wallet is visible inside a vehicle and no person is around, it may tempt a criminal to grasp the opportunity.

\subsubsection{Awareness Theory}
Four components of crime have been suggested in~\cite{brantingham1981notes}: victim, criminal, geo-temporal and legal. Concentrating on the particular spatial aspect of crime is important to understand the behavior of criminals. A crime's space may be chosen either deliberately or accidentally by either the victim or the offender according to their life styles. A number of things have impact on the crime rate of an region. For example, what kind of individuals reside in a particular region and what kind of security can be obtained.

\subsubsection{Broken Windows Theory}
The broken windows theory is a criminal theory of the norm-setting and signaling impact of urban disorder and vandalism on extra offense and anti-social behavior~\cite{wilson1982broken}. The theory claims that sustaining and monitoring urban environment to avoid small offenses like public drinking and vandalism assists to produce an environment of order and lawfulness, thus preventing more severe offenses happening.The theory has been applied as an inspiration for many reforms in criminal policy, such as the controversial mass use of ``stop, question, and frisk" by the New York City Police Department.

\subsubsection{Crime Opportunity Theory}
Crime opportunity theory suggests that analysts should search for concentrations of offense targets. For example, a dense downtown community without off-street parking could have many vehicles parked on the street. This kind of areas might become a region hotspot for thefts from vehicles; while a suburban inhabited by dual-income families will have few persons during weekdays, and their property is unprotected, thus their community can be a region hotspot of burglary. Note that in such condition, many levels of hotspots can exist simultaneously. Within region hotspots, identified by the subdivision in this case, might be streets with increased amounts of burglaries, and some of the houses on these streets might be broken into numerous times.

\subsection{Social Criminal Theories} 
Theories in this family are used in a number of approaches as the conflict theory or structural conflict perspective in sociology are associated with crime. Social criminal theories emphasize poverty, absence of education, lack of marketable abilities, and subcultural values as fundamental causes of crime.

\subsubsection{Social Disorganization Theory}
Social Disorganization Theory directly links crime rates to community environmental characteristics~\cite{shaw1942juvenile}. Social disorganization theory postulates that an individual's residential location is a substantial element shaping the chance the individual will become involved with illegal activities. The theory shows that, among determinants of an individual's later illegal activity, residential location is more substantial than the individual's characteristics (e.g., age, gender, or race). For instance, the theory indicates that youths from disadvantaged community participate in a subculture which approves of delinquency, and these types of youths thus acquire criminality within this social and ethnic setting.

\subsubsection{Social Strain Theory}
Strain theory indicates that mainstream culture is saturated with dreams of opportunity, freedom, and prosperity. The majority of people buy into this dream, and it will become a powerful cultural and psychological inspiration. If the social structure of opportunities is unequal and prevents the majority from realizing the dream, some of the people dejected will use illegal ways (crime) to realize it. Others may retreat or drop out into deviant subcultures (like gang members). Robert Agnew developed this theory to incorporate varieties of strain which were not derived from economic restrictions. This is known as ``General Strain Theory"~\cite{merton1968social}.

\subsubsection{Culture Conflict Theory}
The theory of culture conflict is linked to the disagreement over dissimilarities in values and beliefs. This is based on the idea that different cultures or classes cannot agree on what is common acceptable behavior. For instance, if the upper and middle classes work to make a living in a legal way, others may use illegitimate ways, such as stealing, to make a living.

\subsubsection{Social Efficacy Theory}
Recent evidence points to the role of social efficacy, which is the willingness of nearby residents to intervene with regard to the common good. This is dependent on mutual trust and solidarity among neighbors~\cite{sampson1997neighborhoods}. Communities that have a great deal of social efficacy have less offenses than these at low levels. Social efficacy is not a property of individual people or places, but a characteristic associated with groups of individuals.

\subsubsection{Subcultural Theory}
Subcultural theorists focus on small cultural groups fragmenting aside from the mainstream that form their own beliefs and meanings of life. It indicates that delinquency among lower class youths is a reaction towards the social norms of the middle class~\cite{cohen1955delinquent}. Several youth, especially from poorer regions where opportunities are few, might buy into social norms specific to those places that may consist of ``toughness" and disrespect regarding authority. Criminal acts might result when youths adapt to norms of the deviant subculture~\cite{kornhauser1978social}.

\subsubsection{Control Theory}
Control Theory tries to describe why people do not become offender~\cite{gottfredson1990general}. It recognizes four principal factors: (1) connection to others, (2) belief in ethical validity of principles, (3) responsibility to achievement, and (4) engagement in main-stream activities~\cite{hirschi2002causes}. The more an individual has those factors, the less likely he/she become offender.  This theory is extended with the fact that an individual with low self control is more likely to become offender. 

\subsubsection{Labeling Theory}
Labeling theory claims that when a person is given the label of a offender, they might accept it and continue to commit crime or refuse it. Even those who initially refuse the label can eventually accept it as the label becomes more well-known especially among their peers. This stigma can be much more profound when the labels are about deviancy, and it is believed that this stigmatization can cause deviancy amplification. Klein~\cite{klein1986labeling} did a test which indicated that labeling theory influenced some youth offenders but not others.

\section{Urban Crime Pattern Analysis}
\label{sec:patternsd}
As suggested by theories from criminology, crime is highly related to time and location. Meanwhile, big urban data provides rich information about crime from temporal and spatial perspectives, which cultivated increasing efforts on temporal-spatial pattern analysis. Temporal-spatial pattern analysis is a procedure that obtains understanding from temporal-spatial related sources and generates understanding for crime analysts. In practice, understanding varies among different environments. In order to acquire appropriate relevant kinds of information, various kinds of temporal-spatial pattern analysis techniques should be leveraged~\cite{leong2015review}. 

\subsection{Temporal Pattern Analysis}
Criminal temporal patterns are complicated since temporal resources could be structured in various intervals like weeks, months, seasons, years and others~\cite{leong2015review}. Generally, the temporal crime analysis focuses on learning useful temporal patterns from sequential crime data. Types of temporal pattern analysis can be summarized as follows:
\begin{itemize} 
	\item Crime tendency refers to the change of a type of crime inside a given region and a long-term time period. For example, the property crime rates declined by double-digit percentages from 2008 to 2016 in the USA.  
	\item Crime periodicity is defined as the repeating patterns of crime at time intervals, e.g., seasonal (i.e. annually recurring) crime patterns.
	\item Similarity search of crime aims to search crime sequences that are similar to a given crime sequence.
	\item Sequential behavior analysis tries to find an offender's sequential behaviors before or after committing a crime, e.g., a burglar often buys drug after committing a burglary.
\end{itemize}

\subsection{Spatial Pattern Analysis}
Crimes are not evenly or randomly distributed in an urban area. Typically, crimes are dense in some regions and sparse in others. Spatial pattern analysis aims to learn the aggregation of crime, i.e., hotspots, inside a city. Additionally, crimes are proved to be correlated with environment contexts. In this subsection, we introduce crime hotspots and spatial factor analysis in detail.
\begin{itemize} 
	\item Crime hotspot is defined as a geographic location with more than normal quantity of crime activities, or a location where individuals have greater than normal risk of victimization~\cite{eck2005mapping}. On the contrary, there exist cold-spots with less than the normal density of crime. Some hotspots might be hotter than others because of the difference of crime density. Generally, hotspot analysis finds spatial patterns through spatial clustering. 
	\item Spatial factor analysis aims to find the main spatial factors of crime~\cite{leong2015review}. The major hypothesis of spatial analysis is that crime should correlate with environment contexts and this hypothesis is supported by various criminal theories. For instance, according to routine activity theory, three elements, i.e., a motivated offender, a suitable target or victim, and the absence of a capable guardian, are required to converge in time and space for a crime occurring. 
\end{itemize}

\subsection{Spatio-Temporal Pattern Analysis}
Criminal spatio-temporal pattern analysis aims to obtain understanding from geo- and time-related crime data. The challenge is how to identify patterns from the dynamic interaction among space, time and crime~\cite{leong2015review} since crime patterns are believed to vary with time and location~\cite{skogan1990disorder,loukaitou1999hot,harries1999mapping}. In this subsection, we will review important temporal-spatial patterns of urban crime.
\begin{itemize} 
	\item Earthquake-like pattern: The concentrating patterns suggest that an earthquake is likely to produce a series of aftershocks close to the area of the original earthquake~\cite{d2015statistical}. Similar phenomena are observed in crime formation, such as burglars may repeatedly assault neighboring communities over a time period.  This encourages applying seismology techniques like self-exciting point processes to model urban crime~\cite{daley2007introduction}.
	\item Spatio-temporal hotspot : Crimes such as gang violence happen concentrated in time and space. Spatio-temporal hotspot is defined as a geographic location coupled with a time period where greater than normal amount of crimes occur. It aims to incorporate temporal patterns on spatial hotspot for crime analysis. 
	\item Spatio-temporal correlations: Spatio-temporal correlations are explored in~\cite{zhao2017modeling}. For a urban region, ``intra-region temporal correlation" is observed-- (i) for two consecutive time slots, they are likely to share similar crime numbers; and (ii) with the increase of differences between two time slots, the crime difference has the propensity to increase.  Over all urban regions, ``inter-region spatial correlation" is observed -- (i) two geographically close regions have similar crime numbers; and (ii) with the increase of spatial distance between two regions, the crime difference tends to increase.
\end{itemize}
\section{Computational Tasks for Urban Crime}
\label{sec:task}
Big urban data is typically large-scale, noisy, dynamic and heterogeneous, which calls for efficient and effective computational solutions. Thus, in order to enhance crime research in big urban data era,  numerous computational tasks have been proposed. In this section, we review key computational tasks for urban crime with representative algorithms.

\subsection{Crime Rate Prediction}
Crime rate prediction aims to predict the future crime rate of a given urban region. In this subsection, we categorize crime rate prediction models according to the data they use as prediction based on crime data,  environmental context data, andmsocial media data.

\subsubsection{Prediction Based on Crime Data}
Precise crime prediction 30 days ahead for small areas, like police precincts, is proposed in~\cite{gorr2003short}. Prediction precision of univariate time series models are compared with techniques commonly employed by police. A fixed-effect regression model of absolute percent prediction error suggests that average offense number must be lager than 30 to obtain less than 20\% prediction error.  It is also found that Holt exponential smoothing is the most precise model for precinct-level crime prediction. In~\cite{chen2008forecasting}, autoregressive integrated moving average (ARIMA) is employed for near future prediction of property crime. Based on 50 weeks' property crime data, an ARIMA model is built to predict crime number of 1 week ahead. It is found that ARIMA model has higher fitting and prediction precision than exponential smoothing.  

A four-order tensor for crime forecasting is presented in~\cite{mu2011empirical}. The tensor encodes the longitude, latitude, time, and other related crimes. The tensor can tackle the data sparsity since each order is lower-dimensional. Additionally, the geometry structure is properly maintained in tensor. Leveraging the tensor framework, empirical discriminative tensor analysis algorithm is presented to acquire adequate discriminative information and reduce empirical risk simultaneously. In~\cite{yu2014crime}, a new feature selection and construction method is proposed for crime prediction by using temporal and spatial patterns. Multi-dimensional feature is denoted as spatio-temporal pattern that is built upon regional crime cluster distributions in different levels. Then a Cluster-Confidence-Rate-Boosting framework is presented to combine local spatio-temporal patterns into global crime pattern, which is then employed for crime prediction.

Temporal patterns of dynamics of violence are analyzed using a point process model in the scenario of urban crime prediction~\cite{lewis2012self} . The rate of crimes is partitioned into the sum of a Poisson background rate and a self-exciting component in which crimes induce the growth in the rate of the process. Specifically, each crime produced by the process in turn produces a series of offspring crimes according to a Poisson distribution. The background rate is normally fixed for crimes. In~\cite{mohler2011self},  self-exciting point process models are implemented for predicting crimes.  They leverage a nonparametric evaluation strategy to gain understanding of temporal-spatial triggering function and temporal tendencies in the background rate of burglary. Particularly, spatial heterogeneity in crime rates can be evaluated by background intensity estimation and the self-exciting effects detected in crime data. 

\subsubsection{Prediction Based on Environmental Context Data}
Crime rate tendencies and periodicity are analyzed through a routine activity approach~\cite{cohen1979social,felson1980human} to predict crime.  Specifically, it is assumed that the distribution of events far from homes raises the opportunities for offense and hence yields higher crime rates. The assumption can help understand crime rate tendencies in the United States 1947-1974 as a consequence of changes in such factors as labor force involvement and single-parent families. Seasonal crime patterns were analyzed for urban crime prediction~\cite{field1992effect}. An evaluation of annual, quarterly, and monthly crime data exhibited solid evidence that temperature has a positive impact on most kinds of crime. The influence was independent with seasonal variation. The key explanation is that higher temperatures trigger individuals to spend more time out of home, which is consistent with routine activity explanations for crime and has been revealed to raise the chance of crime. The results indicate that temperature is one major reason to be taken into consideration when describing quarter-to-quarter variations of urban crime. 

\subsubsection{Prediction Based on Social Media Data}
Twitter posts with rich and event-based context is leveraged for predicting criminal incidents~\cite{wang2012spatio}. The framework contains two components. The first component is a spatio-temporal generalized additive model, which leverages a feature-based method to predict future crime at a given location and time. The second component extracts textual information through semantic role labeling-based latent Dirichlet allocation. In addition, a new feature selection approach is designed to discover essential features. A preliminary analysis of Twitter-based crime prediction is proposed in~\cite{wang2012automatic}. The method incorporates intelligent semantic analysis of Twitter posts, and dimensionality reduction through latent Dirichlet allocation. Twitter-specific linguistic analysis and mathematical topic modeling to locate discussion topics of Twitter across a urban are introduced in~\cite{gerber2014predicting}. These topics are integrated into a crime prediction model. The authors find that the addition of Twitter information enhances crime prediction accuracy compared with kernel density estimation. It identifies  several performance bottlenecks that influence the usage of Twitter in a real decision support system. In~\cite{aghababaei2016mining}, Twitter content is employed for crime tendency forecast, in which a Twitter sampling approach is  presented to gather historical data for handling the missing data problem over time. The experiments unveiled the relationship of Twitter content and crime tendency. Besides, some crime types like burglary are found to have closer relationship with the shared Twitter content than other types.

\subsection{Crime Hotspot Detection}
Crime hotspot detection (mapping) is a spatial mapping technique focusing on the identification of the concentration of crime events across the urban. In this subsection, we categorize crime hotspot detection methods based on the types of techniques they leverage, i.e., (1) KDE-based techniques: a non-parametric method to calculate the probability density function of crimes, (2)Reaction-Diffusion-based techniques: a mathematical framework based on reaction-diffusion partial differential equations to learn the dynamic nature of crime hotspots, and (3) Other techniques: they include geographic boundary thematic mapping, grid thematic mapping, spatial ellipses and hotspots optimization tools. 

\subsubsection{KDE-based Techniques}
Prospective crime hotspot detection methods are improved by analyzing interpolation technique, grid cell size, and bandwidth on the prediction precision of Kernel Density Estimation (KDE)~\cite{hart2014kernel}. Particularly, based on variations in essential user-defined settings that are parts of the interpolation process, this work presents scientific explanations of the quality of KDE hotspot maps . The analytical technique contains evaluating these effects across multiple crime types, such as assault, robbery and burglary. To quickly get an accurate hotspot map, an efficient approach to convert KDE hotspot map with low resolution to a new map with contour lines  is designed in~\cite{de2016mskde}. The outcome is a hotspot map with smooth boundaries, which is equally accurate to KDE hotspot maps leveraging smaller cell sizes, but has faster generation speed. The new maps are more natural illustration of the real world's hotspots compared to the original KDE maps. Many helpful suggestions for setting parameters of KDE such as the grid cell size and search radius (bandwidth) are proposed for hotspot detection tasks~\cite{chainey2013gis,eck2005mapping,ratcliffe1999hotbeds}. 

\subsubsection{Reaction-Diffusion-based Techniques}
A computational framework based on reaction-diffusion partially differential equations is developed for studying the formation and dynamics of crime hotspots~\cite{short2010dissipation}. The framework is designed upon empirical evidence for how criminals move and interact with victims. Analysis suggests that crime hotspots from the recurring crimes diffuses locally, but not so far as to mix remote crime together. A nonlinear analysis is proposed to detect the crime hotspots through the reaction-diffusion system~\cite{short2010nonlinear}. The authors discover amplitude equations that control the forming process of crime hotspot patterns through a perturbation method. Different from the super-critical hotspots discovered in existing work~\cite{short2010dissipation},  sub-critical hotspots are discovered that arise upon sub-critical or trans-critical bifurcations with respect to the geometry. Enlightened by~\cite{short2010nonlinear}, a reaction-diffusion based approach is proposed to detect hotspots rigorously~\cite{berestycki2014existence}. More specifically, the existence of steady states is demonstrated with multiple spikes of two types, i.e., (1) Multiple spikes having the same amplitude, and (2) Multiple spikes having different amplitudes. They leverage a strategy according to Liapunov-Schmidt reduction and improve it to the quasilinear crime hotspot model. 

\subsubsection{Other Techniques}
Geographic boundary thematic mapping is a method to represent spatial distributions of urban crimes~\cite{ratcliffe2001crime}, which can rapidly generate hotspot maps and need little knowledge to interpret~\cite{williamson2001tools}. Boundary regions in this method are generally arbitrarily defined by government, e.g., police precinct. Crimes in the hotspot map concentrate on these regions which will then be shaded with respect to crime number inside them. Grid thematic mapping method is developed to mitigate the situation of different sizes and shapes of different regions such as police precincts~\cite{bowers2001gis,lebeau20017}, in which grids with the same size and shape are drawn on a urban map snapping the study area. Thus all regions in the map are of uniform dimensions and are comparable, which is helpful to quickly and easily detect crime hotspots. Spatial ellipses is a hotspot detection software that locates hotspot within the study area~\cite{bates1987spatial}. It first finds the densest aggregation of crime locations on the map (hot clusters), and then fits a ``standard deviation-ellipse" to each one. The ellipses rank crime clusters according to their sizes and properties. An hotspots optimization tool is presented to enhance the detection of hotspots through optimizing its boundary according to the spatial patterns of crime driving factors in~\cite{wang2012optimization,wang2013crime}. A pattern is introduced to indicate a mix of values of related variables which is able to distinguish hotspots and normal regions from the spatial perspective, named Geospatial Discriminative Patterns (GDPatterns)~\cite{ding2009discovery}. The proposed model automatically detects the crime hotspots and identifies GDPatterns between crime hotspots and normal regions simultaneously.  


\subsection{Next-Location Prediction}
Next-Location Prediction aims to predict the location where an offender will commit a crime according to the offender's historical trajectories or other information. A personalized random walk based method for next-location prediction is proposed in~\cite{tayebi2014crimetracer}. To be specific, it leverages co-offending, crime trends and road network data to personalize the random walk process. According to crime pattern theory, criminals usually use their most familiar regions as part of their activity space. A probabilistic model is proposed to capture known criminals' spatial trajectories in their activity space. This setting is then used to predict crime locations for criminals. Correlated walk analysis with several regression routines are designed for forecasting the behavior of a serial criminal in~\cite{levine2006crime}. Based on the analysis of criminal's repeated behavior in time, direction, and distance, the prediction of where and when the next crime will occur is produced. In~\cite{wang2015using}, approaches that incorporate textual content into next-location prediction models are studied based on two hypothesis: (1) An person's future spatial trajectory relates to his/her historical tweets, and (2) Crime rates correlate with the density of users' spatial trajectories in the same region. Additionally, the relationship between these next-location predictions and the incidence of urban crimes is investigated, with the target to help future study into intelligent crime prediction.  In~\cite{qian2011weighted}, Rossmo's formula is introduced to predict crime's next location with geographic profiling techniques. To be specific, a traffic network is incorporated for geographic profiling, and the next-location prediction problem is handled upon weighted traffic network, where shortest path between nodes are leveraged to replace Euclidean distance for accurate next crime location prediction. An agent-based simulation is designed to learn the influence of temporal pulse activities on the criminal location selection process~\cite{fox2012simulating}. The simulation offers a technique to modify spatio-temporal patterns as the consequence of some apriori special activities.

\subsection{Criminal Network Analysis}
Law enforcement and police departments have realized that criminal networks are important for crime analysis and prevention. Typically, a criminal network consists of (1) nodes: the individual actors within the criminal network such as offenders, and (2) ties: the relationships between actors, such as co-offenders and crime gang. In this subsection, we categorize criminal network analysis methods according to the techniques they employ as (1) Agent-based techniques: agents move in network and interact with each other, thus produce complex dynamics and patterns from simple behavioral rules, (2) Graph Theory-based techniques: Metrics (e.g. centrality) and  techniques (e.g. community detection) from graph theory are employed for criminal network analysis, and (3) GIS-based techniques: Geographic Information System techniques are introduced to handle criminal network analysis.

\subsubsection{Agent-based Techniques}
An agent-based model to imitate the formation of street gangs is introduced in~\cite{hegemann2011geographical}. The motion dynamics of agents are combined into an evolving network of street gangs, which are influenced by prior interplays among agents in the system. Datasets of gangs, locations and criminal behaviors are incorporated into the model. The authors discover that highways, rivers, and the center locations of gangs' activities affect the agents' motions. The gang concentration through an interacting agent system is identified on a lattice~\cite{barbaro2013territorial} . Two-gang Hamiltonian framework is presented that agents have red or blue association but are otherwise indistinguishable. In this model, all indirect interplays happen through graffiti marks, on-site and on closest surrounding regions. The dynamic segregation~\cite{schelling1971dynamic} and dynamic subsequent variations~\cite{fossett2006ethnic,macy2006ethnic} models leverage agent-based techniques with structured time-invariable interactions~\cite{perc2013evolutionary} for criminal network analysis. They use structured rather than well-mixed networks since not all actors are connected to others in a gang or criminal network, and the interplays between actors generally follow an fixed manner that will not change over time. 

\subsubsection{Graph Theory-based Techniques}
Analytical Hierarchy Process and  Graph Theory are combined to solve gang crime problem~\cite{gao2014solution}. The key purpose is to find the conspirators and produce a priority list according to message traffic within crime cases. To find one offender, it firstly quantifies the topics through Analytic Hierarchy Process, then builds the network model through Graph Theory and finally proves the relationship between this offender to the identified conspirators and non-conspirators. Criminal gangs are sensed and characterized in networks reconstructed from telephone network data~\cite{ferrara2014detecting}. Specifically, it presents an expert framework to reveal the main structure of criminal networks concealed in telephone data. This framework enables mathematical network analysis and community detection of telephone network data. It allows police departments to deeply understand the structure within criminal networks, discover offender leader and identify relationship among sub-groups. In~\cite{paulo2013social}, several police analytic tasks are handled regarding street gangs from the graph theory perspective. Particularly, it establishes the degree of membership for criminals who do not acknowledge the membership in a street gang, quickly recognizes sets of influential criminals through the tipping model, and decomposes gangs into sub-gangs to identify criminal ecosystems. A unified model is proposed to bridge the conceptual gap between abstract crime dataset and co-offending network in ~\cite{tayebi2011locating}. Then it uses several centrality metrics, such as degree, closeness, betweenness, eigenvector and PageRank centrality, on the co-offending network and studies how leader criminal detection and elimination can help crime prevention. Cluster analysis is applied to detect criminal leaders, gangs and interplays between gangs within criminal network~\cite{xu2003untangling}. More specifically, an concept space idea is introduced to build links among criminals according to the similarities of criminals' crime activities. The more two criminals take part in the same crime activities, the more likely to have a link between two criminals. 

\subsubsection{GIS-based Techniques}
Influence of geography and social networks on gang violence is learned in~\cite{papachristos2013corner}. Applying urban shootings data, it analyzes the impact of geographic proximity, organizational memory and group properties (e.g. reciprocity and transitivity) on gang violence in Chicago and Boston. Results show adjacency of criminal gang turf and previous conflict between gangs are solid predictors of future gang violence. A method for co-offense forecast is proposed in~\cite{tayebi2014spatially}, where geographic, social, geo-social and similarity factors are used to classify offenders. To handle the skewed distribution problem of co-offending networks, it designs three types of criminal cooperation opportunities which are helpful to reduce the class imbalance ratio, while maintaining half of the co-offenses. Criminal network among gang members are discovered in Los Angeles according to the rare observations of a mix of social connections and geographic regions of the individuals ~\cite{van2013community}. A similarity graph is built for the individuals and  spectral clustering is leveraged to identify clusters in the graph. It analyzes various ways to encode the geo-social information on graph structure and the effect on the resulting clusterings. Spatial and social distance are employed to study the connections in youth co-offending network and essential features that influence co-offending network formation. Results demonstrate that spatial distance can better describe the overall structure, and social distance plays a role in network structure of close spatial distance.

\subsection{Near Repeat Victimization}
Crime does not happen randomly or evenly across time or space. The near-repeat victimization identifies the increased risk of repeat victimization at the same or surrounding regions and within a certain time period. This phenomenon has been repeatedly demonstrated for urban crime over the world.  Near-repeat phenomenon is often jointly analyzed with hotspot detection tasks because their similar spatio-temporal distributions, but there are also some other techniques to handle near-repeat victimization, such as social network analysis.

\subsubsection{Hotspots-based Techniques}
Time span of repeat victimization and relationships between repeat victimization, deprivation and burglary hotspots are studied in~\cite{johnson1997new}.  It finds that the time span of repeat victimization presents an exponential manner.  Results also reveal a clear connection between repeat victimization and deprivation, and declare that the geographical area of repeat victimizations may well contribute to the definition of burglary hotspots. Encouraged by the precepts of optimal foraging theory, burglary hotspots are found to shift over time~\cite{johnson2004stability}, such that the repeating phenomenon of hotspots is not predictable over three months, but they have a tendency to move in a slippery way, i.e., moving to surrounding regions at successive time slots. Repeating burglary is studied based on police calls for service data~\cite{townsley2000repeat}. It demonstrates: (1) hotspots can be discovered by mathematical analysis of spatial crime aggregation, (2) unstable hotspots are generally short-term concentrations of hot-dots, while stable hotspots appear to reveal more social and physical features of specific locations. In~\cite{wang2017analysis}, the near-repeat phenomenon is used to analyze the risk levels around hotspots. To be specific, a temporal matrix is proposed to measure the fluctuation of risk around hotspots. The results show that (1) hotspots always exist, (2) space-time regions of high chance are always variable in space and time, (3) locations in the vicinity of hotspots concurrently share higher risk, and (4) crime risks around hotspots follow a wave diffusion process. 

\subsubsection{Other Techniques}
Likelihood distribution functions for the time intervals between repeat crimes are learned in~\cite{short2009measuring}. It compares these distributions to mathematically derived distributions where the repeat effects are due to solely consistent risk heterogeneity. It discovers that a form of event boosts is able to describe the observed distributions, while risk heterogeneity alone cannot, thus it models repeat victimization as a series of random activities, the likelihood of which changes once an crime occurs. Near repeat and social network techniques are incorporated on discovering crimes' spatio-temporal patterns~\cite{hu2017integrating}, in which burglary data of NYC is analyzed and compared upon average clustering coefficient, degree and closeness centrality. The clusters of near repeat crimes are efficiently calculated in~\cite{yin2017taming}. First, R-tree is used to index crimes, where a crime is defined as a node and edges are built by range querying the vertex in R-tree, thus a graph forms. Cohesive subgraph techniques are used to find the crime chains. K-clique, k-truss, k-core plus DBSCAN algorithms are implemented in sequence regarding their diverse range of capacity to locate cohesive subgraphs.

\subsection{Police Patrolling}
Proper patrol route planning is one important application of crime analysis systems, which helps increase the effectiveness of police patrolling and improve public security simultaneously. In this subsection, we categorize this family of models according to the application tasks including (1)Patrol Area Allocation: partitioning a urban to precincts and arranging them to police officers, (2)Patrol Route Planning:  designing the patrolling route for police car, and (3) Other Applications: they include police decision support systems, and spatial reorganization of police agents.

\subsubsection{Patrol Area Allocation}
A novel framework is designed for allocating patrol area against urban crime~\cite{gholami2015extensive,zhang2015keeping,zhang2016keeping}. This framework models the connection between officers and urban crime as a dynamic Bayesian Network (DBN). Next, a series of improvements are made to the basic DBN causing in a compact model that has lower learning error. Further, by analyzing various Markov models, it is found that the number of the crime and the number of the defender in each region can impact the crime forecast, and combining hidden states within DBN can reduce prediction error. A bi-level optimization method are developed to handle the problem of spatial police patrol allocation aiming to enhance crime response speed~\cite{mukhopadhyay2016optimal}. It first designs a linear programming patrol response formulation and then use Bender's decomposition to solve the optimization problem. The major challenge is that offenders may adjust the location and time of crime with respect to police patrols. To deal with this challenge, an iterative Bender's decomposition approach is proposed. In~\cite{zhang2014simulation}, a police patrol discrete-event simulation model is applied to judge the patrol allocation plans. A response surface strategy is developed to discover optimal or sub-optimal allocation plans to improve response speed and reduce workload variation. An iterative searching process is designed to learn the connection between parameters in a redistricting algorithm and performance measures of allocation plans. 

\subsubsection{Patrol Route Planning}
Planning effective and balanced routes for police patrolling is challenging with multiple police cars across different police distinct. Distributed services on road networks are introduced for patrolling route planning~\cite{chen2018balanced}. Specifically, they formulate this problem as a Min-Max Multiple-Depot Rural Postman Problem. To resolve the routing problem, they design an effective tabu-search-based algorithm and present three novel lower bounds to evaluate the routes. Rapid route planning is coupled with interactive spatio-temporal hotspot exploration in~\cite{godwin2017hotsketch} . The major components are: (1) a sketch-based method for dynamic route planning, which allows a police officer to fast establish a route across the city and discover the number and types of crime along the route, and (2) a spatio-temporal hotspot method considering time, location, season, and recent crime volume.  A patrol routes planning method is designed to maximize the coverage of hotspots and minimize the patrolling distance simultaneously~\cite{chawathe2007organizing}. It treats the road network as graph, where node, edges and edge weights refer to intersections, streets and streets' importance. These features as well as the topology of graph are then used to calculate the optimal patrol route. This method allows the automation of patrol planning and the dynamic adjustment in terms of constantly changing environment. A real-time patrol route planning approach for dynamic environments is proposed  in~\cite{chen2010cross,chen2012fast}. It first models the route planning problem in one patrol unit case. Then it designs an efficient algorithm to leverage cross entropy for real-time applications in practice. A graph-based MDP is presented to model the patrol routing process in~\cite{chen2010patrol}, and then it proposes an $\epsilon$-optimal strategy to handle the dimensionality curse problem. This strategy is derived from the idea of $\epsilon$-optimal horizon approximation. 

\subsubsection{Other Applications}
A decision support system, which merges predictive policing abilities with a patrolling districting model, is proposed for the planning of predictive patrolling regions~\cite{camacho2015decision}. The system efficiently and evenly identifies partitions of a district balanced based on the  decision maker's preferences. To analyze the crime records, it designs a method to describe the spatially and temporally indeterminate crime events. A multi-criteria police districting problem is studied with the consideration of the characteristics of region, risk, compactness, and mutual support~\cite{camacho2015multi}. The decision-maker can identify its preferences on the characteristics, workload balance, and efficiency. The model is resolved in the form of a heuristic algorithm. In~\cite{furtado2006using}, an application is proposed for assisting the spatial reorganization of police agents. To model activities of criminals, it employs self-organization strategy where police agents learn from their local regional, and make decisions based on this self-organization strategy and other environmental information. Two models to help police department design policy and prevention strategies against crimes are introduced in~\cite{moonen2007organising}. The first model develops a patrol dispatch strategy where police agents patrol separately without the knowledge of other agents' locations, while in the second strategy, the patrols of police agents are jointly dispatched.

\section{Future Research Directions}
\label{sec:discussion}
In this section, we discuss some insights and present some future directions.

\subsection{More Crime Pattens}
Urban crime is observed to have intrinsically complex spatio-temporal patterns with complicated urban configurations. However, the majority of existing algorithms can capture only certain aspects of patterns. Hence, comprehensive techniques are desired to learn complicated temporal-spatial patterns for urban crime analysis. Recently deep learning has been demonstrated in the capability of capturing complex spatio-temporal patterns for more accurate prediction in many urban computing tasks such as traffic flow forecasting~\cite{zhang2017deep,zhang2016dnn}, and air quality prediction~\cite{qi2017deep}. Therefore, one promising direction is to build novel models such as deep learning models to learn more complex spatio-temporal patterns for advancing urban crime analysis.

\subsection{More Advanced Techniques}
Urban crime is intrinsically complex because of its dynamic interplay with space, time and other factors like economics, environment and urban configuration. Thus more advanced techniques should be introduced and developed to handle crime analysis. For instance, deep reinforcement learning, which can continuously update polices during the interactions with environment, is suitable for capturing the dynamic nature of urban crime data. Besides, crime in urban is impacted by multiple sources such as meteorological data, point of interests (POIs) data and human mobility data~\cite{zhao2017modeling}. The majority of existing algorithms handle multiple sources equally or in a linear manner, but fail to capture the nonlinear connections and subordinations among multiple sources~\cite{kang2017prediction}. This calls for advanced techniques to effectively incorporate features from multiple sources for crime analysis. Additionally, feature extraction from multiple sources should follow an automatic way since hand-crafted features are insufficient to capture complex spatio-temporal patterns, which encourages to leverage End-to-End frameworks to automatically combine feature extraction and computational tasks of urban crime.

\subsection{More Computational Tasks}
The recent development of big data techniques has greatly advanced urban crime analysis, which provides unprecedented and unique opportunities to designs more sophisticated models to tackle practical policing tasks in real world. As an example, the stop-question-and-frisk program in New York City is a crime prevention policy that temporarily detains, questions and searchs citizens on the street for weapons and contraband. However, this practice has been complained about its racism and failure to reduce  crime such as burglary and robbery. Thus more efforts should be made on crime prevention strategies that maximize deterrent value and minimize infringement on the rights of citizens simultaneously. Besides, multiple urban tasks should be jointly considered for a safer and smarter city such as  education, health, urban planning, economic development, employment, police, justice, immigration, poverty, integration, etc.

\subsection{Urban Simulation}
Policing strategies must be pre-evaluated before their active use in order to save the unnecessary cost of deployment and avoid negative impacts on urban safety. Thus, a urban  environment simulation is necessary for the offline evaluation and visualization of new policing strategies. Moreover, urban environment simulation can allow researchers and police department to investigate, gain insights, and develop policing techniques to boost their communities, integrating the interplays between crimes, economy, and urban configurations.

\section{Conclusion}
\label{sec:conclusion}
Crime analysis plays a tremendously impactful role in the sustainable development of urban and the quality of citizen's life. With recent advances in urban data sensing, collecting and  integrating technologies, a large amount of fine-grained urban crime-related data has been recorded with rich environmental and social information, which motivates a variety of computational tasks to advance urban crime analysis. In this article,  we give an overview about urban crime from a computational perspective.  We first review two families of criminal theories, i.e., environmental criminal theories and social criminal theories, and then key crime patterns in mining urban crime data.  We introduce major computational urban crime tasks with representative algorithms. We also discuss some interesting research directions about computational crime with big urban data. 

\section*{Acknowledgements}
This research is supported by the National Science Foundation (NSF) under grant number IIS-1714741 and IIS-1715940.

\bibliographystyle{abbrv}
\bibliography{crime}  
\end{document}